\documentclass[usenatbib]{emulateapj}

\newcommand{\aanda}{{\it A\&A}}
\newcommand{\be}{\begin{equation}}
\newcommand{\ee}{\end{equation}}

\newcommand{\ltorder}{\hbox{ \rlap{\raise 0.425ex\hbox{$<$}}\lower
0.65ex\hbox{$\sim$} }}
\newcommand{\gtorder}{\hbox{ \rlap{\raise 0.425ex\hbox{$>$}}\lower
0.65ex\hbox{$\sim$} }}

\shorttitle{IMBHS in Early Globular Clusters}
\shortauthors{E.Vesperini, S.L.W. McMillan, A. D'Ercole, F. D'Antona}

\begin{document}

\title{Intermediate-Mass Black Holes in Early Globular Clusters}

\author{Enrico Vesperini, Stephen L.W. McMillan}
\affil{Department of Physics, Drexel University, Philadelphia, PA 19104}
\author{Annibale D'Ercole}
\affil{INAF, Osservatorio Astronomico di Bologna, via Ranzani 1,
  I-40127 Bologna, Italy}
\author{Francesca D'Antona}
\affil{INAF, Osservatorio Astronomico di Roma, via di Frascati 33,
  I-00040 Monteporzio, Italy}
\begin{abstract}
  Spectroscopic and photometric observations show that many globular
  clusters host multiple stellar populations, challenging the common
  paradigm that globular clusters are ``simple stellar populations''
  composed of stars of uniform age and chemical composition.  The
  chemical abundances of second-generation (SG) stars constrain the
  sources of gas out of which these stars must have formed, indicating
  that the gas must contain matter processed through the
  high-temperature CNO cycle.  First-generation  massive
  Asymptotic Giant Branch (AGB) stars have been proposed as the source
  of this gas. In a previous study, by means of hydrodynamical and
  N-body simulations, we have shown that the AGB ejecta collect in a
  cooling flow in the cluster core, where the gas reaches high
  densities, ultimately forming a centrally concentrated subsystem of
  SG stars. In this Letter we show that the high gas density can also lead to
significant accretion onto a pre-existing seed black hole. We show
that gas accretion can increase the black hole mass by up to a
factor of 100.  The details of the gas dynamics are important in
determining the actual black hole growth.  Assuming a near-universal
seed black hole mass and small cluster-to-cluster variations in the
duration of the SG formation phase, the outcome of our scenario is one
in which the present intermediate-mass black hole (IMBH) mass may have
only a weak dependence on  
the current cluster properties.  The scenario presented provides a
natural mechanism for the formation of an IMBH
at the cluster center during the SG star-formation phase.

\end{abstract}
\keywords{globular clusters: general}
\section{Introduction}
\label{sec:intro}
If globular clusters follow the same ``$M_{BH}-\sigma$'' relation
between central black hole mass and effective velocity dispersion as
is observed in galaxies (Tremaine et al. 2002), many clusters should
host intermediate-mass black holes (IMBHs) with masses in the range
$10^2 \ltorder M_{BH}/M_{\odot} \ltorder 10^4$.  Unfortunately, hard
observational evidence for IMBHs in globular clusters has proved
elusive.  Currently the strongest case is the massive cluster G1 in
M31, where dynamical, X-ray, and radio studies are consistent with a
mass $M_{BH}\sim2\times10^4~$ (Gebhardt et al. 2005, Ulvestad et
al. 2007). In our own Galaxy, a measurement of an IMBH mass $M_{BH}
\sim 4\times10^4 M_{\odot}$ has been reported for $\omega$ Centauri
(Noyola et al. 2008), although this result has been questioned in a
recent study (Anderson \& van der Marel 2010, van der Marel \&
Anderson 2010) suggesting that the IMBH signature becomes much weaker
when a more accurate estimate of the cluster center is used.  Recently
Ibata et al. (2009) have reported evidence of a density and kinematic
cusp in the core of M54, a cluster located at the center of the
Sagittarius dwarf galaxy; their models suggest that the cusp could be
due to a $\sim 10^4 M_{\odot}$ IMBH.
 
For most Galactic globular clusters, kinematic and structural data
currently provide only upper limits on possible IMBH masses
(McLaughlin et al. 2006, Pasquato et al. 2009, van der Marel \&
Anderson 2009). However, these upper limits are larger than the values
of $M_{BH}$ expected from the $M_{BH}-\sigma$ relation, and do not
exclude the possibility that many globular clusters host IMBHs with
masses in the above range (van der Marel \& Anderson 2009).

The presence of an IMBH in a globular cluster has important
consequences for the cluster's structure, kinematics, internal
energetics and long-term dynamical evolution (see e.g. Baumgardt et
al. 2004, Heggie et al. 2007, Trenti et al. 2007, Miocchi 2007; 
see also McMillan 2008 for a review).  In
addition, Eddington accretion onto IMBHs has been proposed as the
mechanism powering some ultraluminous X-ray sources (e.g. Farrell et
al. 2009), and IMBHs have long been recognized as potential rich
sources of gravitational waves detectable by LISA (Miller 2009). For
these reasons, despite the lack of strong observational constraints,
the formation and consequent properties of IMBHs have become a topic of
considerable theoretical and observational interest.

One possible IMBH formation mechanism is direct collapse of a very
massive ($\geq 250 M_{\odot}$) Population III star (see e.g. Bond et
al. 1984, Madau \& Rees 2001 and references therein; but see Abel et
al. 2002 for hydrodynamical simulations suggesting that these stars
would form in isolation).
For non-primordial or more massive IMBHs, the leading formation
mechanisms center on dynamical processes in dense star clusters.
These include runaway mergers of massive stars in clusters with high
central densities (Portegies Zwart \& McMillan 2002, G\"urkan et
al. 2004, Portegies Zwart et al. 2004), and merging of black holes in
binaries (see O'Leary et al. 2006 and references therein).  However,
numerous potential problems affecting these models have been pointed
out in the recent literature, suggesting that the net growth rate and
hence the final mass of the resulting IMBH may be much lower than
suggested by earlier dynamical simulations (Yungelson et al. 2008,
Glebbeek et al. 2009).

Thus, while there appear to be numerous plausible stellar, binary, and
dynamical pathways to the formation of relatively ``low-mass''
($\sim100-200 M_{\odot}$) IMBHs in clusters, there is currently no
clear consensus on a mechanism for the production of high-mass
($10^3-10^4 M_{\odot}$) IMBHs.

In this Letter we propose a new scenario for the growth of IMBHs in
early globular clusters.  It is a natural consequence of the physical
conditions that may have obtained in cluster cores during the
formation of the second generation (hereafter SG) stars now observed
in many globular clusters.  Specifically, we show that accretion by a
pre-existing seed black hole of just a small fraction of the gas from
which the SG stars formed can lead to black holes in the high-mass
($10^3-10^4 M_{\odot}$) IMBH range.

 The structure of this Letter is as follows.
In Section \ref{sec:multip} we briefly review the observational
evidence for multiple stellar populations in globular clusters, and
summarize the main elements of our model for the formation and
evolution of multiple populations.  In Section \ref{sec:bh}, we
present our scenario for IMBH growth.  We discuss some consequences of
this scenario in Section \ref{sec:concl}.

\section{Multiple stellar populations in globular clusters}
\label{sec:multip}
Spectroscopic and photometric observations provide strong evidence
for multiple stellar generations in globular clusters.

Spectroscopic studies have shown that, even in clusters whose stars
have a common Fe abundance, there are substantial star-to-star
differences in the abundances of lighter elements (see Gratton et
al. 2004 and references therein, and Carretta et al. 2009a,b for a
recent spectroscopic study of 19 Galactic clusters).  All the Galactic
globular clusters for which spectroscopic studies have been carried
out show evidence of multiple stellar populations, and in all cases SG
stars make up a significant fraction ($50-80\%$) of the total
(D'Antona \& Caloi 2008, Carretta et al. 2009a,b).

Several recent photometric studies have provided further support for
these results by showing the photometric fingerprints of multiple
populations (multiple main sequences, significant main sequence
broadening, or subgiant branch splits) in the color-magnitude diagrams
both of ``normal'' clusters in the Galaxy (Piotto et al. 2007, Marino
et al. 2008, Milone et al. 2008, Anderson et al. 2009), and of more
complex systems such as $\omega$ Cen (Lee et al. 1999, Pancino et
al. 2000, Bedin et al. 2004, Villanova et al. 2007), M22 (Marino et
al. 2009, Lee et al. 2009) and probably Terzan 5 (Ferraro et
al. 2009).

The differences in light-element abundances suggest that SG stars
formed out of gas containing matter processed through high-temperature
CNO cycle reactions in first-generation (FG) stars.  The main
candidates currently suggested as possible sources of gas for SG
formation are massive AGB stars (Ventura et al. 2001) and rapidly
rotating massive stars (Decressin et al. 2007). Although many aspects
of these two scenarios are still under investigation, in order to form
the large mass of SG stars suggested by observations, both scenarios
require that either the IMF of the FG stars was highly anomalous, with
an unusually large fraction of massive stars, or the FG population had
a normal IMF but was initially at least ten times more massive than is
now observed.

We have recently carried out hydrodynamical and N-body simulations to
explore the formation and dynamical evolution of multiple population
clusters formed out of AGB star ejecta (D'Ercole et al. 2008).  Our
simulations show that the gas lost by FG AGB stars collects in a
cooling flow in the cluster core, where most SG stars are formed.  We
have further demonstrated a dynamical mechanism whereby, after the
centrally concentrated SG stars have formed, the
cluster's response to early mass loss (of intracluster gas and/or
supernova ejecta) leads to the escape of a substantial fraction of FG
stars.  During the subsequent long-term evolution, the SG and FG
populations mix by two-body relaxation, plausibly leading to the
multiple-population clusters observed today.

We now explore the implications for the growth of a pre-existing seed
black hole of the physical conditions in the central regions of the
cluster during the SG formation phase.

\section{Growth of a seed black hole}
\label{sec:bh}

In the hydrodynamical simulations presented in D'Ercole et al. (2008),
we assumed that the cluster FG stars followed a King (1962) density
profile
\begin{equation}
    \rho=\rho_0 \left[1+\left({r \over r_c}\right)^2\right]^{-3/2}
\end{equation} 
out to a truncation radius $R_t$, with stellar masses distributed in
the range $0.1<m/m_{\odot}<100$ according to a ``Kroupa'' initial mass function (Kroupa
et al. 1993).  We adopted a concentration 
$c=\log(R_t/r_c)=1.5$ and followed the evolution of the AGB ejecta in
clusters having different values of the total FG mass $M$ and
truncation radius $R_t$.

Figure 1 shows the time evolution of the AGB ejecta density and sound
speed within 0.1 pc of the cluster center during the SG formation
phase, for four clusters spanning broad ranges in initial mass and
tidal radius.  In all cases, if the cluster already hosts a seed black
hole with mass $M_{BH} \sim 10^2~M_{\odot}$---the result of one of the
``prompt'' low-mass IMBH formation mechanisms discussed previously,
operating in the FG population---the gas conditions shown in Figure 1
will lead to accretion.  The Bondi accretion rate onto a black hole of
mass $M_{BH}$ is 
\begin{eqnarray}
\dot M &\simeq& 3 \times 10^{-4} \left(M_{BH}\over 100
       M_{\odot}\right)^2\left(\rho\over 10^{-18} \hbox{g
       cm}^{-3}\right) \nonumber\\
       &~&~~~~~~~~~~~~~~~~~~~~~
       \times \left({3 \hbox{km s}^{-1}\over c_s}\right)^3
       ~\hbox{M}_{\odot}\,\hbox{yr}^{-1}
\end{eqnarray}
where the density $\rho$ and the sound speed $c_s$ have been
normalized to the typical values found in our simulations and shown in
Figure 1. For the physical conditions in our four simulations, the Bondi
accretion rate is $\sim 10^{-4}-10^{-3} ~\hbox{M}_{\odot}/\hbox{yr}$,
much larger than the Eddington rate,$~~$
$\sim3\times10^{-9}\,(M_{BH}/M_{\odot})/\eta\,\,\hbox{M}_{\odot}$/yr (Krolik
1999), where $\eta$ is the radiative efficiency, to be discussed
further below.  The black hole will therefore accrete gas at the
Eddington rate.

\begin{figure}
\begin{center}
 \includegraphics[width=3.5in]{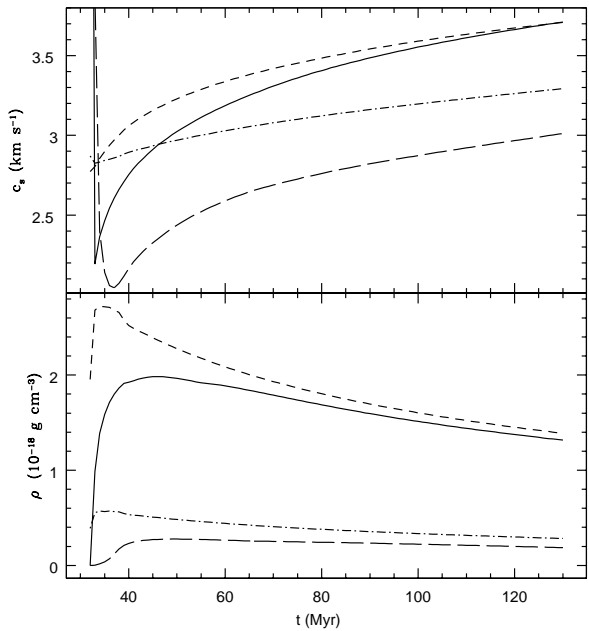} 
 \caption{Time evolution of the AGB ejecta density (lower panel) and
   sound speed (upper panel) within 0.1 pc of the cluster center
   during the SG formation phase for a cluster with FG total mass $M$
   and tidal radius $R_t$ equal to $(10^7~M_{\odot}, 200~pc)$ (solid
   line), $(10^7~M_{\odot}, 40~pc)$ (dashed line), $(10^6~M_{\odot},
   90~pc)$ (long-dashed line), $(10^6~M_{\odot}, 18~pc)$ (dot-dashed
   line).}
\end{center}
\end{figure}

Under these circumstances, the black hole mass will grow on a Salpeter
time scale
\begin{equation}
    T_S \simeq 3.9\times 10^8 \,\eta\, L_E/L ~\hbox{yr},
\end{equation}
where $L_E \sim 1.45 \times 10^{38} M_{BH}/M_{\odot}$ erg/s is the
Eddington luminosity and a value of the mass per electron $\mu_e\sim
1.15$ has been adopted (see e.g. Krolik 1999).  The value of $\eta$ is
uncertain and depends on the properties of the accretion disk and on
the spin of the accreting black hole.  Here we adopt $\eta\sim
0.05-0.1$, a range commonly adopted in studies of black hole growth in
galactic nuclei.  We note that values of $\eta$ larger than 0.1,
appropriate for rapidly rotating black holes, would reduce the growth
of the IMBH, while much smaller values ($\eta \ll 1$ ) typical of
thick accretion disks would significantly increase the IMBH growth
rate.  For accretion at the Eddington rate with $\eta\sim 0.05-0.1$,
$T_S\sim 20-40$ Myr.

The duration of the SG star formation episode in the models of
D'Ercole et al. (2008) is $\Delta T \approx 10^2$ Myr.  This time
scale is constrained by spectroscopic observations showing little
variation in the total CNO of FG and SG stars, implying that only AGB
progenitors with $M>4-5 M_{\odot}$ must have contributed gas to the SG
formation process.  Thus, if the seed black hole accretes at close to
the Eddington rate during this time, its mass will grow by a factor of
$\sim 10-100$, leading, by the end of the SG formation phase, to a
final black hole mass of $M_{BH} \sim 10^3-10^4~M_{\odot}$.
$~~$\\

\section{Discussion}
\label{sec:concl}

Several aspects of the scenario just presented merit further
discussion and exploration.

We consider first the effect on the cooling flow of the energy
radiated by the accreting black hole. 
To address this issue we follow
the analysis of black hole outflow presented by King (2003, 2005), in
which it is shown that the wind produced by the black hole sweeps up
the surrounding gas into an expanding shell.  Because of effective
 energy losses due to inverse Compton
scattering, the shell expands in the momentum-conserving regime.
Under these conditions (see King
2003, 2005), in order for the momentum flux from a black hole accreting at
the Eddington rate to overcome gravity and expel the shell from the
cluster, the black hole mass must be larger than
\begin{equation}
    M_{BH,crit} = \frac{f_g\kappa\sigma^4}{\pi G^2}.
\end{equation}
Here, $\kappa$ is the electron scattering opacity, $f_g$ is the
fraction of the total cluster mass in the form of gas, and it has been
assumed that the gas is embedded in an isothermal system with velocity
dispersion $\sigma$.

Assuming $\sigma \gtorder 20~\hbox{km/s}$ (the smallest value for
which the AGB ejecta can be retained) and $f_g\approx0.05$ (D'Ercole
et al. 2008), we find $M_{BH,crit} \gtorder 10^4 M_{\odot}$, much
larger than the assumed mass of our initial seed black hole.  Thus
feedback from the accreting black hole is {\em not} sufficient to
alter significantly the global dynamics of the cooling flow or the SG
star formation process.

On the other hand, the effects of feedback may significantly alter the
local gas dynamics in the vicinity of the black hole.  For example,
recent detailed 2-D hydrodynamical models of accretion onto a 100
$M_{\odot}$ seed black hole in a dense protogalactic cloud
(Milosavljevic et al. 2009) show intermittent accretion, reducing the
net accretion rate to about $1/3$ of Eddington.  Such a reduction in
the mean accretion rate in our case would reduce the growth of the
seed black hole to a factor $\sim2-5$ for the range of $\eta$ values
adopted in \S\ref{sec:bh}.

We note here that during the accretion phase the black hole will be
very luminous and might be observed as a (possibly intermittent)
ultraluminous X-ray source (ULX).  We caution however that, before
linking this scenario to the ULXs observed in nearby young massive
clusters (see e.g. Portegies Zwart et al. 2010), one must first verify
whether these clusters (1) meet the conditions to form a seed black
hole, (2) are sufficiently massive and (3) have the structural properties
required for SG star formation.

A second important issue is the possible relation between the IMBH
mass and the current structural properties of the parent cluster.

Any correlation between the mass of the seed black hole and the early
properties of the cluster should be preserved during the Eddington
growth phase, provided that the duration $\Delta T$ of the SG star
formation episode is approximately independent of the cluster
environment. The mass of the seed black hole depends on its formation
mechanism.  If it is the result of stellar evolution in a massive
Pop. III progenitor, a roughly universal mass seems the most likely
result (see e.g. Madau \& Rees 2001). 
Early dynamical simulations of the runaway merger scenario
suggested that the mass of the runaway should scale with the mass of
the cluster (or the cluster core) in which it forms (G\"urkan et
al. 2004, Portegies Zwart et al. 2004).  However, as mentioned in
\S\ref{sec:intro}, strong stellar winds may well limit the actual mass
attained, possibly to as little as a few hundred solar masses, again
largely independent of the initial cluster mass.

For the black hole growth phase to occur, the cluster must {\em
  initially} have been massive and/or compact enough to retain the AGB
ejecta that subsequently flowed into the cluster core and formed the
SG population.  As discussed in D'Ercole et al. (2008), multiple
population clusters must have undergone an early phase of strong FG
mass loss and structural evolution; it is therefore not
straightforward to connect the initial properties of a cluster during
the growth of the seed black hole with the current properties of the
cluster hosting an IMBH.  However, given the loss of most of the FG
population and the possible near-universality of the seed black hole
mass, the most likely outcome of the scenario described here seems to
be one in which the present IMBH mass has only a weak dependence on
current cluster properties.
$~~~~$\\
{\bf Acknowledgments.}
EV and SM were supported in part by NASA grants NNX07AG95G,
NNX08AH15G, and NNX10AD86G and by NSF grant AST-0708299.
AD acknowledges financial support from italian MIUR through grant PRIN 2007
(prot. 2007JJC53X). FD was  supported by PRIN MIUR 2007 'Multiple
Stellar Populations in Globular Clusters: Census, Characterization
and Origin' (prot. n. 20075TP5K9).

\end{document}